\documentstyle[aps,epsfig,floats]{revtex}

\tighten
\newcommand{\beq}{\begin{equation}}
\newcommand{\eeq}{\end{equation}}
\newcommand{\ba}{\begin{eqnarray}}
\newcommand{\ea}{\end{eqnarray}}
\newcommand{\nn}{\nonumber}

\newcommand{\psibar}{\overline{\psi}}
\newcommand{\la}{\langle}
\newcommand{\ra}{\rangle}
\newcommand{\amp}[1]{\la #1 \ra}

\newcommand{\g}{\gamma}
\newcommand{\Z}{{\scriptscriptstyle Z}}
\newcommand{\W}{{\scriptscriptstyle W}}
\newcommand{\sT}{{\scriptscriptstyle T}}
\newcommand{\lra}{\leftrightarrow}
\newcommand{\Pslash}{\kern 0.2 em P\kern -0.56em \raisebox{0.3ex}{/}}

\newcommand{\bm}[1]{\bbox{#1}}
\newcommand{\simorder}{\raisebox{-4pt}{$\, \stackrel{\textstyle >}{\sim} \,$}}

\begin{document}
 
\draft
\title{
\begin{flushright}
\begin{minipage}{4 cm}
\small
hep-ph/9907504\\
RIKEN-BNL preprint\\
VUTH 99-19\\
FNT/T-99/14
\end{minipage}
\end{flushright}
Angular dependences in electroweak semi-inclusive leptoproduction}

\author{Dani\"el Boer$^1$, R. Jakob$^2$ and P.J. Mulders$^3$}
\address{\mbox{}\\
$^1$RIKEN-BNL Research Center\\
Brookhaven National Laboratory, Upton, New York 11973, U.S.A.\\
\mbox{}\\
$^2$Universit\`{a} di Pavia and INFN, Sezione di Pavia\\ 
Via Bassi 6, 27100 Pavia, Italy\\
\mbox{}\\
$^3$Department of Physics and Astronomy, Free University \\
De Boelelaan 1081, NL-1081 HV Amsterdam, the Netherlands
}

\maketitle
\begin{center}\today \end{center}

\begin{abstract}
We present the leading order unpolarized and polarized cross sections in 
electroweak semi-inclusive deep inelastic leptoproduction. The azimuthal 
dependences in the cross section differential in the transverse momentum of 
the vector boson arise due to intrinsic transverse momenta of the quarks. 
However, the presented asymmetries are not suppressed by inverse powers of 
the hard scale. 
We discuss the different opportunities to measure specific asymmetries as 
offered by neutral compared to charged current processes and point out the
optimal kinematical regions. The present and (proposed) future HERA collider 
experiments would be most suitable for measuring some of the asymmetries 
discussed here, especially in case of $\Lambda$ production.  
\end{abstract}


In this article we extend results of Refs.\
\cite{Mulders-Tangerman-96,Boer1,Jakob-98}  
on semi-inclusive deep inelastic leptoproduction cross sections to include 
contributions from $Z$-boson exchange and $\gamma$-$Z$ interference terms in
neutral current processes as well as
contributions from $W$-boson exchange in charged current processes. 
The azimuthal 
dependences in the cross section differential in the transverse momentum of 
the vector boson arise due to intrinsic transverse momenta of the quarks.
Only leading order $(1/Q)^0$ effects are 
discussed, since higher twist contributions like those of Refs.\
\cite{Cahn,Berger,Liang} are expected to be negligible at
energies for which electroweak contributions are relevant. 
Also, we will focus on tree 
level, i.e., order $(\alpha_s)^0$. A rich structure nevertheless arises when
taking into account the polarization of the initial or final state particles. 
At the end of this article we will discuss the experimental opportunities to
study specific terms in the cross sections. 

For details of the calculation and the formalism we refer to 
\cite{Mulders-Tangerman-96,Boer1}. We shortly repeat the essentials. 
It is convenient to use the hadron momenta in the process 
$\ell H \rightarrow \ell^\prime h X$ to define two lightlike 
vectors $n_+$ and $n_-$, satisfying $n_+\cdot n_-$ = 1. 
These vectors then define the lightcone components of a vector as $
p^\pm \equiv p\cdot n_\mp$ and we use the component notation
$p=[p^-,p^+,\bm{p}_\sT^{}]$. Up to mass terms the momentum $P$ of the target
hadron ($H$) is along $n_+$, the momentum $P_h$ of the outgoing 
hadron ($h$) is along $n_-$.
We assume here that we are discussing current quark fragmentation, for which
one requires $P\cdot P_h \sim Q^2$, where $q^2 = - Q^2$ is the
momentum transfer squared. In leading order in $1/Q$ the process
factorizes into a product of a hard perturbative partonic subprocess and 
two soft
nonperturbative parts, which describe the distribution of quarks in the
target and the final fragmentation of the struck quark into 
hadrons, respectively.  

The neutral current cross section for unpolarized and polarized electroweak
semi-inclusive lepton-hadron scattering is given by 
\begin{equation}
\frac{d\sigma(\ell H\to \ell' h X)}
     {d\Omega\, dx\, dz\, d^{\,2}{\bm q_\sT^{}}}= 
\frac{\alpha^2 z\; y}{8\; Q^4} \sum_{ij} L_{\mu\nu}^{ij}\, 2M\, 
{\cal W}_{i j}^{\mu\nu} \, \chi_{ij} \;.
\label{cross2}
\end{equation}
We use invariants $x=Q^2/(2\,P\cdot q)$, $z=P \cdot P_h/P\cdot q$
and $y= (P \cdot q)/ (P \cdot l) \approx q^-/l^-$ 
($l$ being the momentum of the beam 
lepton). The cross section is differential in $dx$, $dz$,
$d\Omega = 2\,dy\,d\phi$, and 
in $d^{\,2}{\bm q_\sT^{}}$ where 
$q_\sT= q +x\,P - P_h/z = [0,0,\bm q_\sT^{}]$.  
The indices $i,j$ can be $\g$ for the photon or $Z$ for the 
$Z$-boson. The relative propagator factors $\chi_{ij}$ are given by
\begin{equation}
\chi_{\g\g}  = 1 \;,\qquad
\chi_{\g\Z}  =  \chi_{\Z\g} = 
\frac{1}{\sin^2 (2 \theta_W)} \, \frac{Q^2}{Q^2+M_Z^2} \;,\qquad
\chi_{\Z\Z}  =  \left(\chi_{\g \Z}\right)^2 \;. 
\end{equation} 
Here we note that in this process the scale $Q$ is defined by the {\em 
spacelike\/} 
vector boson momentum $q$ (with $Q^2 \equiv -q^2$), hence the width of the
$Z$-boson plays a
negligible role and we have taken it to be zero. 
Also, we will consider the cross section differential in the transverse
momentum of the vector boson, but the factorized expression \cite{Collins-93b} 
that we will consider 
will require $|\bm{q}_\sT|^2 \ll Q^2$, to insure that one is sensitive to the
region of intrinsic transverse momenta. 
 
The lepton tensor (neglecting the lepton masses) is given by 
\begin{equation} 
L_{\mu\nu}^{ij} (l, l^\prime)
=  C^{ij}\, \left[ 2 l_\mu l^\prime_\nu
+ 2 l_\nu l^\prime_\mu - Q^2 g_{\mu\nu} \right]+ D^{ij} \, 
2i \,\epsilon_{\mu\nu\rho\sigma} l^\rho l^{\prime \sigma} \;,
\label{leptten2}
\end{equation}
where the incoming lepton has momentum $l$ and the back-scattered lepton
momentum $l'$.
We have defined 
\ba
&& C^{\g\g}=1, \quad C^{\g Z}=C^{Z\g}=e^l (g_V^l-g_A^l\lambda_e), 
\quad C^{ZZ}= (g_V^l{}^2 + g_A^l{}^2)-(2 g_V^l g_A^l)\lambda_e \;,\\
&& D^{\g\g}=\lambda_e, \quad D^{\g Z}=D^{Z\g}=e^l (g_V^l\lambda_e-g_A^l), 
\quad D^{ZZ}=(g_V^l{}^2 + g_A^l{}^2)\lambda_e-(2 g_V^l g_A^l) \;,
\ea 
where $e^l$ denotes the coupling of the photon to the leptons in units of 
the positron charge; $g_V^l$, $g_A^l$ denote the vector and axial-vector
couplings of the $Z$-boson to the leptons, respectively and $\lambda_e$ is the
helicity of the incoming lepton. 

To leading order 
the expression for the hadron tensor, including quarks and anti-quarks, is
\begin{eqnarray}
2M\, {\cal W}_{i j}^{\mu\nu}&=& \int dp^- dk^+ d^{\,2}\bm{p}_\sT^{} 
d^{\,2} \bm{k}_\sT^{}\, 
\delta^2(\bm{p}_\sT^{}+\bm{q}_\sT^{}-\bm{k}_\sT^{})\, 
\left. \text{Tr}\left( \Phi (p) 
V_i^\mu \Delta 
(k) V_j^\nu \right)\right|_{p^+, \,k^-} + \left(\begin{array}{c} 
q\leftrightarrow -q \\ \mu \leftrightarrow \nu
\end{array} \right) \;,
\label{Wmunustart}
\end{eqnarray}
where $p$ and $k$ represent the quark momentum before and after the
interaction with the vector boson. 
The vertices $V_i^\mu$ can be either the photon vertex 
$V_\g^\mu = e \gamma^\mu$ or the
$Z$-boson vertex $V_\Z^\mu = g_V \gamma^\mu + g_A \gamma_5 \gamma^\mu$.
The vector and axial-vector couplings to the $Z$ boson are given by:
\begin{eqnarray} 
g_V^k &=& T_3^k - 2 \, e^k\,\sin^2 \theta_W \;,\\
g_A^k &=& T_3^k \;,
\end{eqnarray} 
where $e^k$ denotes the charge and $T_3^k$ the weak isospin of 
particle $k$ (i.e., $T_3^k=+1/2$ for $k=e^+,\mu^+,\nu,u$ and 
$T_3^k=-1/2$ for $k=e^-,\mu^-,\bar \nu,d,s$). 
We have omitted flavor indices and summation. The 
correlation functions $\Phi$ and $\Delta$ comprise information on 
the hadronic structure of the target in terms of quark degrees of 
freedom and on the quark hadronization process, respectively. They 
are given by (path-ordered exponentials are suppressed):
\begin{eqnarray}
\Phi_{mn}(P,S;p)&=& \int \frac{d^{\,4\!}x}{(2\pi)^4}\ e^{ip \cdot x}\,
\amp{P,S|\psibar_n (0) \psi_m (x) | P,S} \;,\\
\Delta_{mn}(P_h,S_h;k) & = & 
\sum_X \int \frac{d^{\,4\!}x}{(2\pi)^4}\ e^{ik\cdot x}\,
\langle 0 \vert \psi_m(x) \vert P_h,S_h; X \rangle
\langle P_h, S_h; X \vert \overline \psi_n(0) \vert 0 \rangle\;.
\end{eqnarray}
Using Lorentz invariance, hermiticity, and parity invariance, the 
(partly integrated) correlation function 
\begin{equation} 
\Phi(x,\bm{p}_\sT) \equiv  \left. \int dp^-\ \Phi(P,S;p) 
\right|_{p^+ = x P^+, \bm{p}_{\scriptscriptstyle T}}
=\Phi^{(O)}(x,\bm{p}_\sT)+\Phi^{(L)}(x,\bm{p}_\sT)+\Phi^{(T)}(x,\bm{p}_\sT)\;,
\end{equation} 
(with indices $O,L,T$ indicating the polarization of the target: unpolarized,
longitudinally and transversely polarized, respectively)
is parametrized in terms of distribution functions as:
\begin{eqnarray} 
\Phi^{(O)}(x,\bm{p}_\sT) &=& 
\frac{M}{2P^+}\,\Biggl\{
     f_1(x,\bm{p}_\sT^2)\,\frac{\Pslash}{M}+
{}+ h_1^\perp(x,\bm{p}_\sT^2)\,\frac{\sigma_{\mu\nu} p_\sT^\mu P^\nu}{M^2}
\Biggr\}\;,
\nonumber \\ 
\Phi^{(L)}(x,\bm{p}_\sT) &=& 
\frac{M}{2P^+}\,\Biggl\{
{}- \lambda\,g_{1L}(x,\bm{p}_\sT^2) 
    \frac{\Pslash \gamma_5}{M}
{}- \lambda\,h_{1L}^\perp(x,\bm{p}_\sT^2) 
    \frac{i\sigma_{\mu\nu}\gamma_5 p_\sT^\mu P^\nu}{M^2} 
\Biggr\}\;,
\nonumber \\ 
\Phi^{(T)}(x,\bm{p}_\sT) &=& 
\frac{M}{2P^+}\,\Biggl\{
{}  f_{1T}^\perp(x,\bm{p}_\sT^2)\, \epsilon_{\mu \nu \rho \sigma}\gamma^\mu 
	\frac{P^\nu p_\sT^\rho S_{\sT}^\sigma}{M^2}
{}- \frac{\bm{p}_\sT\cdot \bm S_\sT}{M}\,
	g_{1T}(x,\bm{p}_\sT^2)
     \frac{\Pslash \gamma_5}{M}
\nonumber\\ && \hspace{20mm}
{}- h_{1T}(x,\bm{p}_\sT^2)\,
     \frac{i\sigma_{\mu\nu}\gamma_5 S_{\sT}^\mu P^\nu}{M}
{}- \frac{\bm{p}_\sT\cdot \bm S_\sT}{M}\,h_{1T}^\perp(x,\bm{p}_\sT^2)
    \frac{i\sigma_{\mu\nu}\gamma_5 p_\sT^\mu P^\nu}{M^2}
\Biggr\}\;,
\end{eqnarray} 
with $M$ being the target hadron mass. We only consider polarization 
of spin-1/2 hadrons, represented by $\lambda = M\,S^+/P^+$ the 
lightcone helicity and 
$\bm S_\sT$ the transverse spin of the target hadron.
The normalization is chosen by requiring that 
$\int dx \, d^{\,2} \bm{p}_\sT \, f_1^a(x,\bm{p}_\sT^2) = n^a$,
where $n^a$ is the number of valence quarks with flavor a. 
Time reversal invariance is expected to eliminate the T-odd functions
$f_{1T}^\perp$ and $h_1^\perp$, especially in the case of semi-inclusive deep
inelastic scattering due to the absence of initial state interactions,
cf.\ Refs.\ \cite{Collins-93b,Anselmino-Leader,Boer1}. Nevertheless, there is a
possibility that they might be effectively generated by a gluonic background
field, cf.\ e.g.\ \cite{Boer2}, 
hence we keep these functions for completeness. 

The (partly integrated) correlation function 
\begin{equation} 
\Delta(z,\bm{k}_\sT) \equiv  \left. \frac{1}{z} \int dk^+\ \Delta(P_h,S_h;k)
\right|_{k^- = P_h^-/z,\ \bm{k}_{\scriptscriptstyle T}} =
\Delta^{(O)}(z,\bm{k}_\sT)+\Delta^{(L)}(z,\bm{k}_\sT)
+\Delta^{(T)}(z,\bm{k}_\sT)\;,
\end{equation} 
(now $O,L,T$ indicating the polarization of the observed final state hadron) 
is parametrized in terms of fragmentation functions as:
\begin{eqnarray}
\Delta^{(O)}(z,\bm{k}_\sT)&=&
\frac{M_h}{P_h^-} \Biggl\{
D_1(z,z^2\bm{k}_\sT^2)\,\frac{\Pslash_h}{M_h}
{}+ H_{1}^\perp(z,z^2\bm{k}_\sT^2)\,
   \frac{\sigma_{\mu \nu} k_\sT^\mu P_h^\nu}{M_h^2}
\Biggr\}\;,
\nonumber \\ 
\Delta^{(L)}(z,\bm{k}_\sT)&=&
\frac{M_h}{P_h^-} \Biggl\{
{}- \lambda_h\,G_{1L}(z,z^2\bm{k}_\sT^2)\,\frac{\Pslash_h \gamma_5}{M_h}
{}- \lambda_h\,H_{1L}^\perp(z,z^2\bm{k}_\sT^2)\,
\frac{i\sigma_{\mu\nu}\gamma_5\, k_\sT^\mu P_h^\nu}{M_h^2}
\Biggr\}\;,
\nonumber \\ 
\Delta^{(T)}(z,\bm{k}_\sT)&=&
\frac{M_h}{P_h^-} \Biggl\{
D_{1T}^\perp(z,z^2\bm{k}_\sT^2)\, \frac{\epsilon_{\mu \nu \rho \sigma}
\gamma^\mu P_h^\nu k_\sT^\rho S_{h\sT}^\sigma}{M_h^2}
{}- \frac{(\bm{k}_\sT^{}\cdot \bm{S}_{hT}^{})}{M_h}\, G_{1T}(z,z^2\bm{k}_\sT^2)
  \,\frac{\Pslash_h \gamma_5}{M_h}
\nonumber\\ && \hspace{20mm}
{}- H_{1T}(z,z^2\bm{k}_\sT^2)\,
    \frac{i\sigma_{\mu\nu}\gamma_5\,S_{h\sT}^\mu P_h^\nu}{M_h}
{}- \frac{(\bm{k}_\sT^{}\cdot \bm{S}_{h\sT}^{})}{M_h}\, 
    H_{1T}^\perp(z,z^2\bm{k}_\sT^2)\,
\frac{i\sigma_{\mu\nu}\gamma_5\, k_\sT^\mu P_h^\nu}{M_h^2}
\Biggr\}\;,
\label{Deltaexp}
\end{eqnarray}
with $M_h$ the mass, $\lambda_h = M_h\,S_h^-/P_h^-$ the lightcone helicity, 
and $\bm S_{h\sT}$ the transverse spin of the produced spin-1/2 hadron.
The
choice of factors in the definition of fragmentation functions is such that
$\sum_a\int dz\,z^2\, d^{\,2}\bm{k}_\sT\,D_1^a(z,z^2\bm{k}_\sT^2) = N_h$, 
where $N_h$ is the total number of produced hadrons. The fragmentation   
functions $H_1^\perp$ and $D_{1T}^\perp$ are called T-odd functions, 
which in contrast to the T-odd distribution functions {\em are\/} 
expected to be present, since they are not forbidden by
time reversal invariance due to the presence of final state 
interactions, cf.\ Ref.\ \cite{Collins-93b}. 

We would like to emphasize that the distribution and fragmentation functions
as defined above parametrize the soft nonperturbative parts of the scattering
process. At present a complete calculation of these nonperturbative 
objects from first principles, as for instance by a lattice calculation, is 
not 
available. Estimates can be made using QCD sum rule or model calculations.
Apart from the known functions $f_1(x)$, $g_1(x)$ and $D_1(z)$, 
which are functions integrated over the transverse momentum (there exists also
marginal information on $G_1(z)$ \cite{DeFlorian}), recently first 
experimental hints on the size of some asymmetries concerning 
the so-called Collins function $H_1^\perp$ \cite{Collins-93b} have been 
presented \cite{H1perp}. 
None of the other functions has been 
experimentally accessed. Therefore, many 
of the observables that we will discuss contain unknown functions, which 
however are not just parametrizations of ignorance, but represent essential 
information on the structure of hadrons as can be probed in hard scattering 
processes and on the hadronization process. For the interpretation of 
the various functions we refer to Refs.\ \cite{Mulders-Tangerman-96,Boer1}. 

In processes with at least two hadrons one needs to be careful with
the notion of {\em transverse}. 
In the definition of distribution and fragmentation functions (which one wants
to be independent of the specifics of the process),
transverse momentum components denoted by a subscript T are defined with
respect to the momenta $P$ and $P_h$ such that $\bm{P}_\sT^{} =0$ and
$\bm{P}_{h\sT}^{} = 0$, respectively. Consequently vectors are decomposed
in plus, minus and transverse components defined by the lightlike
vectors $n_+$ and $n_-$, constructed from $P$ and $P_h$, and
the transverse projector 
$g_\sT^{\mu\nu} \equiv \ g^{\mu\nu}- n_+^{\,\{\mu} n_-^{\nu\}}$.  

\begin{figure}[htb]
\begin{center}
\psfig{file=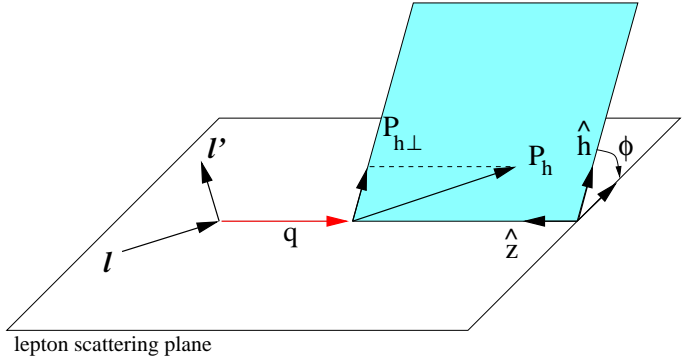, width=9cm}
\vspace{0.2 cm}
\caption{\label{fig1} Kinematics for one-particle inclusive leptoproduction.
The lepton scattering plane is determined by the momenta
$l$, $l^\prime$ and $P$. The azimuthal out-of-plane angle $\phi$ of the 
produced hadron is indicated.}
\vspace{-5 mm}
\end{center}
\end{figure}

For defining the (process dependent) 
azimuthal angles with respect to the scattering 
plane, on the other hand, it is more convenient to use a frame where the 
virtual boson and the target, i.e.~the momenta $q$ and $P$, are 
collinear. We indicate transverse components in the latter frame 
with the subscript $\perp$ (and call them {\em perpendicular} 
henceforth). Thus depending on the choice of frame the covariantly
defined vector $q_\sT^{\,\mu} = q^{\,\mu} +x\,P^{\,\mu} - P_h^{\,\mu}/z$ 
can be  the transverse component of $q$ with respect to the collinear 
hadrons, or it is up to a factor
$1/z$ the component of $P_h$ perpendicular to the scattering 
plane, $q_\sT = -P_{h\perp}/z$. The kinematics in the 
frame where $q$ and $P$ are collinear can be expressed
by the set of normalized vectors:
\ba
\hat t &\equiv & \frac{2x}{Q} \tilde{P} \;,\\
\hat z &\equiv & -q/Q \;,\\[1 mm]
\hat h &\equiv & -q_\sT/Q_\sT = -(q+x\, P - P_h/z)/Q_\sT \;,
\ea
where $Q_\sT^2=-q_\sT^2$ and 
$\tilde{P} \equiv P-(P\cdot q) \, q/q^2$, such that:
\begin{eqnarray}
n_+^\mu & = & 
\frac{1}{\sqrt{2}} \left[ \hat t^\mu + \hat z^\mu \right]\;, 
\label{nplusc}
\\
n_-^\mu & = & 
\frac{1}{\sqrt{2}} \left[ \hat t^\mu - \hat z^\mu
+2\,\frac{Q_\sT^{}}{Q}\,\hat h^\mu \right]\;.  
\label{nplusc2}
\end{eqnarray}
The lepton momentum reads
\begin{equation}
l^\mu = \frac{2-y}{2y} \, \hat t^\mu -\frac{Q}{2}\,  \hat z^\mu +
Q\frac{\sqrt{1-y}}{y} \, \hat l_\perp^\mu \;,
\end{equation} 
and we define the tensors
\begin{eqnarray}
&& g_{\perp}^{\mu\nu}
\equiv  g^{\mu\nu} -\hat t^\mu \hat t^\nu +
\hat z^\mu \hat z^\nu \;, \\
&& \epsilon_\perp^{\mu \nu} \equiv
-\epsilon^{\mu \nu \rho \sigma} \hat t_\rho \hat z_\sigma \;.
\end{eqnarray}
The cross sections are obtained from the hadron tensor after contraction with
the lepton tensor Eq.\ (\ref{leptten2}), which to leading order in $1/Q$ is 
\begin{eqnarray}
L_{ij}^{\mu \nu} & = & C_{ij} \,\frac{Q^2}{y^2}\, \left[
- 2\,A(y)\,g_\perp^{\mu \nu}
+4\,B(y)\,
\left( \hat l_\perp^\mu\hat l_\perp^\nu +\frac{1}{2}\,g_\perp^{\mu \nu}
\right) \right]- D_{ij} \,\frac{Q^2}{y^2}\, 
i\,C(y)\,\epsilon_\perp^{\mu \nu} \;,
\end{eqnarray}
expressed in terms of the functions
\ba
A(y) &=& \left(1-y+\frac{1}{2}y^2\right) \;, \\
B(y) &=& (1-y) \;,\\[2 mm]
C(y) &=& y (2-y) \;.
\ea
Azimuthal angles inside the perpendicular plane are defined with respect 
to $\hat l_\perp^\mu$, defined to be the normalized perpendicular part of the 
incoming lepton momentum $l$,
\begin{eqnarray}
\hat{l}_\perp \cdot a_\perp &=& - |{\bm a}_\perp | \cos \phi_a \;,
\label{anglecos} \\
\epsilon^{\mu\nu}_\perp \hat{l}_{\perp\mu} a_{\perp\nu}&=& |{\bm a}_\perp | 
\sin \phi_a \;,
\label{anglesin}
\end{eqnarray}
for a generic vector $a$. In the cross sections we will encounter dependences
on the three azimuthal angles $\phi$, $\phi_S$ and $\phi_{S_h}$ of the
produced hadron momentum, its spin vector, and of the target
hadron spin vector, respectively (cf.~Fig.~\ref{fig1}).
We would like to note that at leading order the spin 
vector $\bm{S}_\sT^{}$ is identical to the spin vector 
perpendicular to 
$\hat z$, i.e.\ $\bm{S}_\perp^{}$, and 
also $\lambda = M \, (S\cdot \hat z)/ (P\cdot \hat z)$ (and analogously 
for $\bm{S}_{h\sT}^{}$ and $\lambda_h$). This does 
not hold at order $1/Q$, cf.\ Ref.\ \cite{Boer3}. 

In order to present our results on cross sections in a compact form
we define the following combinations of the couplings and $Z$-boson 
propagators 
\ba
K_1^a(y)&=& A(y)
\left[ C^{\g\g} e_a^2 \chi_{\g\g}
    + 2C^{\g\Z} e_a g_V^a \chi_{\g\Z} 
    + C^{\Z\Z} c_1^a \chi_{\Z\Z}
\right] 
- \frac{C(y)}{2} \left[2D^{\g\Z} e_a g_A^a \chi_{\g\Z} 
                     +  D^{\Z\Z} c_3^a \chi_{\Z\Z} 
\right] \;,
\label{eq:defK2}
\\
K_2^a(y)&=& -A(y)
\left[2C^{\g\Z} e_a g_A^a \chi_{\g\Z} + C^{\Z\Z} c_3^a \chi_{\Z\Z}
\right] 
+\frac{C(y)}{2}\left[ D^{\g\g}e_a^2 \chi_{\g\g} 
                    +2D^{\g\Z} e_a g_V^a \chi_{\g\Z} 
                    + D^{\Z\Z} c_1^a \chi_{\Z\Z} \right],\\[1 mm] 
K_3^a(y)&=& -B(y)
\left[ C^{\g\g} e_a^2 \chi_{\g\g}
    + 2C^{\g\Z}e_a g_V^a \chi_{\g\Z} 
    + C^{\Z\Z} c_2^a \chi_{\Z\Z}
\right] \;, 
\ea
which contain the combinations of the Z-boson-to-quark couplings
\begin{eqnarray} 
c_1^a &=&\left(g_V^a{}^2 + g_A^a{}^2 \right) \;,
\\[2 mm]
c_2^a &=&\left(g_V^a{}^2 - g_A^a{}^2 \right) \;,
\\[3 mm]
c_3^a &=&2 g_V^a g_A^a \;.
\end{eqnarray}
Furthermore, we use the convolution notation (where $w$ denotes 
a weight function)
\begin{equation} 
{\cal F}\left[w\left(\bm{p}_\sT^{},\bm{k}_\sT^{}\right) f\, D\right]\equiv \;
\int d^{\,2}\bm{p}_\sT^{}\; d^{\,2}\bm{k}_\sT^{}\;
\delta^2 (\bm{p}_\sT^{}+\bm q_\sT^{}-\bm 
k_\sT^{}) \, w\left(\bm{p}_\sT^{},\bm{k}_\sT^{}\right)  f^a(x,\bm{p}_\sT^2) 
D^a(z,z^2\bm{k}_\sT^2) \;.
\end{equation}
 
We find for the leading order {\em unpolarized\/} cross section, taking into
account both photon and $Z$-boson contributions,
\beq
\frac{d\sigma(\ell H\to \ell' h X)}{d\Omega dx dz d^{\,2}{\bm q_\sT^{}}}=
\frac{\alpha^2\, x\, z^2\, s}{Q^4}\;\sum_{a,\bar a} \;\Bigg\{ 
          K_1^a(y)\;{\cal F}\left[f_1 D_1\right]+K_3^a(y)\cos(2\phi)\;
             {\cal F}\left[\left(2\,\bm{\hat h}\!\cdot \!
\bm{p}_\sT^{}\,\,\bm{\hat h}\!\cdot \! \bm{k}_\sT^{}\,
                    -\,\bm{p}_\sT^{}\!\cdot \! \bm{k}_\sT^{}\,\right)
                    \frac{h_1^{\perp} H_1^{\perp}}{M_1M_2}\right]
\Bigg\} \;,
\label{LO-OOO}
\eeq
where the sum runs over all quark (and anti-quark) flavors. Perturbative QCD
corrections to the term proportional to $f_1 D_1$, which is independent of the
azimuthal angle at tree level, will also produce 
a $\cos(2\phi)$ term, next to a $\cos(\phi)$ term and the time-reversal odd 
$\sin(\phi)$ and $\sin(2\phi)$ terms, but all these will be
suppressed by inverse powers of the hard scale
\cite{Cahn,Berger,Liang,Hagiwara,Ahmed}.
In order to differentiate between the perturbatively generated and the above 
given $\cos(2 \phi)$ asymmetry, one could for instance apply a transverse
momentum cut-off to exclude the contributions from intrinsic transverse
momentum \cite{Chay} or one can study the 
analogous charged current exchange process, since 
the $\cos(2 \phi)$ term of Eq.\ (\ref{LO-OOO}) will then be absent ($K_3=0$) 
as we will observe below.        

In case the {\em target hadron is polarized\/} the cross section 
is found to be
\begin{eqnarray} 
\lefteqn{
\frac{d\sigma(\ell \vec{H} \to \ell' h X)}{d\Omega dx dz 
d^{\,2}{\bm q_\sT^{}}}=
\frac{\alpha^2\, x\, z^2\, s}{Q^4}\;\sum_{a,\bar a} \;\Bigg\{ 
\lambda\;K_2^a(y)\;\;
         {\cal F}\left[g_{1} D_1\right]}
\nonumber\\ && 
       \qquad - {\lambda}\; K_3^a(y)\sin(2\phi)\;
             {\cal F}\left[\left(2\,\bm{\hat h}\!\cdot \!
\bm{p}_\sT^{}\,\,\bm{\hat h}\!\cdot \!
\bm{k}_\sT^{}\,
                    -\,\bm{p}_\sT^{}\!\cdot \!
\bm{k}_\sT^{}\,\right)
                    \frac{h_{1L}^{\perp}H_1^{\perp}}{MM_h}\right]
\nonumber\\ &&
       \qquad + |\bm S_{T}^{}|\;K_1^a(y)\;\sin(\phi-\phi_{S})\; 
             {\cal F}\left[\,\bm{\hat h}\!\cdot \!\bm{p}_\sT^{}\,
                    \frac{f_{1T}^{\perp}D_1}{M}\right]
              + |\bm{S}_\sT|\;K_2^a(y)\;\cos(\phi-\phi_{S})\;
         {\cal F}\left[\bm{\hat h}\!\cdot\!\bm{p}^{}_\sT\,
             \frac{g_{1T} D_1}{M}\right]
\nonumber\\ && 
       \qquad - |\bm S_{T}^{}|\;K_3^a(y)\sin(\phi+\phi_{S})\;
             {\cal F}\left[\,\bm{\hat h}\!\cdot \!\bm{k}_\sT^{}\,
                    \frac{h_1H_1^{\perp}}{M_h}\right]
\nonumber\\ && 
        \qquad - |\bm S_{T}^{}|\;K_3^a(y)\sin(3\phi-\phi_{S}) 
\;  {\cal F}\left[\left(
              4\,(\!\,
\bm{\hat h}\!\cdot \!\bm{p}_\sT^{}\,\!)^2\,\bm{\hat h}\!\cdot \!\bm{k}_\sT^{}
              -2\,\bm{\hat h}\!\cdot \!\bm{p}_\sT^{}\,\,
\bm{p}_\sT^{}\!\cdot \!\bm{k}_\sT^{}\,
              -\,
\bm{p}_\sT^2\,\bm{\hat h}\!\cdot \!\bm{k}_\sT^{}\, \right)
                 \frac{h_{1T}^{\perp} H_1^{\perp}}{2M{}^2M_h}\right]
\Bigg\} \;,
\label{reducedO}
\end{eqnarray}
where we have not included the unpolarized cross section terms again, i.e.\ 
parts which cancel from differences of cross sections with reversed
polarization are suppressed here and in the following.

The cross section for an {\em unpolarized target, but with spin vector
of the final state (spin-1/2) hadron\/} being determined is
\begin{eqnarray} 
\lefteqn{
\frac{d\sigma(\ell H \to \ell' \vec{h} X)}{d\Omega dx dz 
d^{\,2}{\bm q_\sT^{}}}=
\frac{\alpha^2\, x\, z\, s}{Q^4}\;\sum_{a,\bar a} \;\Bigg\{ 
\lambda_h\;K_2^a(y)\;\;
         {\cal F}\left[f_{1} G_1\right]}
\nonumber\\ && 
       \qquad + {\lambda_h}\; K_3^a(y)\sin(2\phi)\;
             {\cal F}\left[\left(2\,\bm{\hat h}\!\cdot \!
\bm{p}_\sT^{}\,\,\bm{\hat h}\!\cdot \!
\bm{k}_\sT^{}\,
                    -\,\bm{p}_\sT^{}\!\cdot \!
\bm{k}_\sT^{}\,\right)
                    \frac{h_1^{\perp}H_{1L}^{\perp}}{MM_h}\right]
\nonumber\\ &&
       \qquad - |\bm S_{h\sT}^{}|\;K_1^a(y)\;
\sin(\phi-\phi_{S_h})\; 
             {\cal F}\left[\,\bm{\hat h}\!\cdot \!\bm{k}_\sT^{}\,
                    \frac{f_1D_{1T}^{\perp}}{M_h}\right]
              + |\bm{S}_{h\sT}|\;K_2^a(y)\;
         \cos(\phi-\phi_{S_h})\;
         {\cal F}\left[\bm{\hat h}\!\cdot\!\bm{k}^{}_\sT\,
             \frac{f_1 G_{1T}}{M_h}\right]
\nonumber\\ && 
       \qquad + |\bm S_{h\sT}^{}|\;K_3^a(y)\sin(\phi+\phi_{S_h})\;
             {\cal F}\left[\,\bm{\hat h}\!\cdot \!\bm{p}_\sT^{}\,
                    \frac{h_1^{\perp}H_1}{M}\right]
\nonumber\\ &&
        \qquad + |\bm S_{h\sT}^{}|\;K_3^a(y)\sin(3\phi-\phi_{S_h}) 
\;  {\cal F}\left[\left(
              4\,(\!\,
\bm{\hat h}\!\cdot \!\bm{k}_\sT^{}\,\!)^2\,\bm{\hat h}\!\cdot \!\bm{p}_\sT^{}
              -2\,\bm{\hat h}\!\cdot \!\bm{k}_\sT^{}\,\,
\bm{k}_\sT^{}\!\cdot \!\bm{p}_\sT^{}\,
              -\,
\bm{k}_\sT^2\,\bm{\hat h}\!\cdot \!\bm{p}_\sT^{}\, \right)
                 \frac{h_{1}^{\perp} H_{1T}^{\perp}}{2M{}^2M_h}\right]
\Bigg\} \;.
\label{pol-h}
\end{eqnarray} 
The Eqs.~(\ref{reducedO}) and (\ref{pol-h}) show some similarity; the latter
is obtained from the former by the set of replacements 
\{$\text{distribution functions} \lra \text{fragmentation functions}$, 
$M\lra M_h$, $k\lra p$, $\lambda\to\lambda_h$,
$S_{\sT}\to S_{h\sT}$, $\phi_{S}\to\phi_{Sh}$\} together with an additional 
minus sign for each time-reversal odd function 
$f_{1T}^\perp, h_1^\perp, D_{1T}^\perp, H_1^\perp$, where 
the replacement of the distribution functions
by the fragmentation functions means that $f,g,h$ functions are replaced by
$D,G,H$ functions, respectively (and vice versa). 

Finally, the leading order {\em double polarized\/} cross section is 
found to be
\begin{eqnarray} 
\lefteqn{
\frac{d\sigma(\ell \vec{H}\to \ell' \vec{h} X)}
    {d\Omega dx dz d^{\,2}{\bm q_\sT^{}}}=
\frac{\alpha^2\, x\, z^2\, s}{Q^4}\;\sum_{a,\bar a} \;\Bigg\{ 
\lambda\;\lambda_h\;\frac{K_1^a(y)}{2} \;
           {\cal F}\left[g_{1} G_{1}\right]}
\nonumber\\ && \qquad 
        + \lambda\;\lambda_h\;\frac{K_3^a(y)}{2}\;\cos(2\phi)\; 
           {\cal F}\left[\left(
                      2\,\bm{\hat h}\!\cdot\!\bm{p}^{}_\sT\,
                       \,\bm{\hat h}\!\cdot\!\bm{k}^{}_\sT\,
                      -\,\bm{p}^{}_\sT\!\cdot\!\bm{k}^{}_\sT\,\right)
               \frac{h_{1L}^{\perp} H_{1L}^{\perp}}{MM_h} \right]       
\nonumber\\ && \qquad 
      + \lambda\;|\bm{S}_{h\sT}|\;K_1^a(y)\;\cos(\phi-\phi_{S_h})\;
           {\cal F}\left[\bm{\hat h}\!\cdot\!\bm{k}^{}_\sT\,
               \frac{g_{1} G_{1T}}{M_h}\right]
\nonumber\\ && \qquad 
      - \lambda\;|\bm S_{h\sT}^{}|\;K_2^a(y)\;
                \sin(\phi-\phi_{S_h})\;
         {\cal F}\left[\,\bm{\hat h}\!\cdot \!\bm{k}_\sT^{}\,
                \frac{g_1 D_{1T}^\perp}{M_h}\right]
\nonumber\\ && \qquad 
      + \lambda\;|\bm{S}_{h\sT}|\;K_3^a(y)\;\cos(\phi+\phi_{S_h})\;
           {\cal F}\left[\bm{\hat h}\!\cdot\!\bm{p}^{}_\sT\, 
               \frac{h_{1L}^{\perp} H_1}{M}\right]
\nonumber\\ && \qquad 
      + \lambda\;|\bm{S}_{h\sT}|\;K_3^a(y)\;\cos(3\phi-\phi_{S_h})\;
           {\cal F}\left[\left(
                      4\,\bm{\hat h}\!\cdot\!\bm{p}^{}_\sT\,
                     (\,\bm{\hat h}\!\cdot\!\bm{k}^{}_\sT\,)^2
                     -2\,\bm{\hat h}\!\cdot\!\bm{k}^{}_\sT\,
                       \,\bm{p}^{}_\sT\!\cdot\!\bm{k}^{}_\sT\,
                     -\,\bm{\hat h}\!\cdot\!\bm{p}^{}_\sT\,
                      \bm{k}_\sT^2\right)
               \frac{h_{1L}^{\perp} H_{1T}^{\perp}}{2MM_h^2}\right] 
\nonumber\\ && \qquad  
       + |\bm S_{T}^{}|\;
                        |\bm S_{h\sT}^{}|\;\frac{K_1^a(y)}{2}\;
               \cos(2\phi-\phi_{S}-\phi_{S_h})\;
            {\cal F}\left[\,\bm{\hat h}\!\cdot \!\bm{p}_\sT^{}\,
                   \,\bm{\hat h}\!\cdot \!\bm{k}_\sT^{}\,
                   \frac{f_{1T}^\perp D_{1T}^\perp
                        +g_{1T} G_{1T}}{MM_h}\right]
\nonumber\\ && \qquad 
       - |\bm S_{T}^{}|\;
                         |\bm S_{h\sT}^{}|\;\frac{K_1^a(y)}{2}\;
               \cos(\phi-\phi_{S})\;\cos(\phi-\phi_{S_h})\;
            {\cal F}\left[\,\bm{p}_\sT^{}\!\cdot \!
                     \bm{k}_\sT^{}\,
                   \frac{f_{1T}^\perp D_{1T}^\perp}{MM_h}\right]
\nonumber\\ && \qquad 
       + |\bm S_{T}^{}|\;
                         |\bm S_{h\sT}^{}|\;\frac{K_1^a(y)}{2}\;
               \sin(\phi-\phi_{S})\;\sin(\phi-\phi_{S_h})\;
            {\cal F}\left[\,\bm{p}_\sT^{}\!\cdot \!
                     \bm{k}_\sT^{}\,
                   \frac{g_{1T} G_{1T}}{MM_h}\right]
\nonumber\\ && \qquad 
    +|\bm S_{T}^{}|\;|\bm S_{h\sT}^{}|\;K_2^a(y)\;
         \sin(2\phi-\phi_{S}-\phi_{S_h})\;
         {\cal F}\left[\,\bm{\hat h}\!\cdot \!\bm{p}_\sT^{}\,
                \,\bm{\hat h}\!\cdot \!\bm{k}_\sT^{}\,
                \frac{f_{1T}^\perp G_{1T}}{MM_h}\right]
\nonumber\\ && \qquad 
    -|\bm S_{T}^{}|\;
                                |\bm S_{h\sT}^{}|\;K_2^a(y)\;
         \cos(\phi-\phi_{S})\sin(\phi-\phi_{S_h})\;
         {\cal F}\left[\,\bm{p}_\sT^{}\!\cdot \!
                  \bm{k}_\sT^{}\,
                \frac{f_{1T}^\perp G_{1T}}{MM_h}\right]
\nonumber\\ && \qquad 
      + |\bm{S}_{T}|\;|\bm{S}_{h\sT}|\;\frac{K_3^a(y)}{2}
            \cos(\phi_{S}+\phi_{S_h})\;
           {\cal F}\left[h_1 H_1\right]
\nonumber\\ && \qquad 
      + |\bm{S}_{T}|\;|\bm{S}_{h\sT}|\;K_3^a(y)\;
            \cos(2\phi+\phi_{S}-\phi_{S_h})\;
           {\cal F}\left[\left(2(\bm{\hat h}\!\cdot\!\bm{k}^{}_\sT)^2
                               -\bm{k}_\sT^2\right)
               \frac{h_1 H_{1T}^{\perp}}{2M_h^2}\right]
\nonumber\\ && \qquad 
      + |\bm{S}_{T}|\;|\bm{S}_{h\sT}|\; \frac{K_3^a(y)}{2}
            \cos(4\phi-\phi_{S}-\phi_{S_h})\;
           {\cal F}\Bigg[\left(8(\bm{\hat h}\!\cdot\!\bm{p}^{}_\sT)^2
                    (\bm{\hat h}\!\cdot\!\bm{k}^{}_\sT)^2\right.
\nonumber\\ &&\hspace{13mm}\left.
                  -4\,\bm{\hat h}\!\cdot\!\bm{p}^{}_\sT\,
                    \,\bm{\hat h}\!\cdot\!\bm{k}^{}_\sT\,
                    \,\bm{p}^{}_\sT\!\cdot\!\bm{k}^{}_\sT\,
                  -2(\bm{\hat h}\!\cdot\!\bm{k}^{}_\sT)^2\bm{p}_\sT^2
                  -2(\bm{\hat h}\!\cdot\!\bm{p}^{}_\sT)^2\bm{k}_\sT^2
                  +\bm{p}_\sT^2\bm{k}_\sT^2\right)
                \frac{h_{1T}^{\perp} H_{1T}^{\perp}}{4M^2M_h^2}\Bigg]
   \quad  +\quad {\;1\longleftrightarrow 2\;\choose 
             \;\bm{p}\longleftrightarrow\bm{k}\;} \quad
\Bigg\} \;, \label{pol-Hh}
\end{eqnarray}
where the additional term indicated by the parentheses in the last line 
stands for the set of replacements
\{$\text{distribution functions} \lra \text{fragmentation functions}$, 
$M\lra M_h$, $k\lra p$, $\lambda\lra\lambda_h$,
$S_{T}\lra S_{h\sT}$, $\phi_{S}\lra\phi_{Sh}$\} together with an additional 
minus sign for each time-reversal odd function 
$f_{1T}^\perp, h_1^\perp, D_{1T}^\perp, H_1^\perp$.

We remark that the appearance of the couplings in the cross 
sections (\ref{LO-OOO})-(\ref{pol-Hh}) shows a clear pattern. All 
convolutions with chiral-odd distribution functions 
$h_1^\perp, h_{1L}^\perp, h_{1T}, h_{1T}^\perp$ and 
fragmentation functions 
$H_1^\perp, H_{1L}^\perp, H_{1T}, H_{1T}^\perp$ couple 
with $K_3^a$. The chiral-even sector involves the couplings $K_1^a$ 
and $K_2^a$: In Eqs.~(\ref{LO-OOO}) and (\ref{pol-Hh}) the convolutions 
with an even (odd) number of time-reversal odd functions couple 
with $K_1^a$ ($K_2^a$); in the single polarized cross sections 
(\ref{reducedO}) and (\ref{pol-h}) the situation is 
reversed, i.e.\ convolutions 
with an even (odd) number of time-reversal odd functions couple 
with $K_2^a$ ($K_1^a$).

\begin{figure}[htb]
\begin{center}
\psfig{file=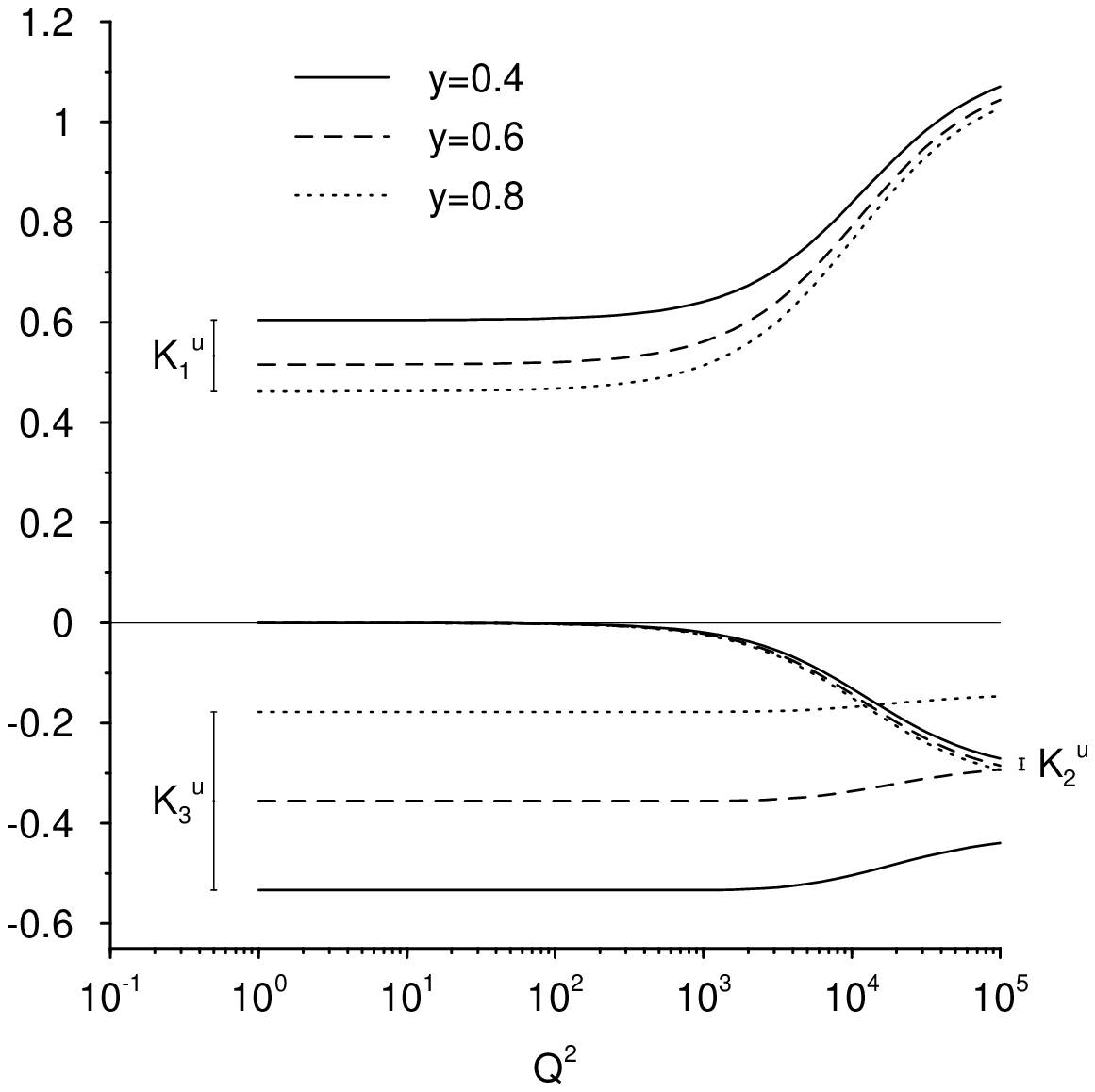, width=8.2cm}\quad
\psfig{file=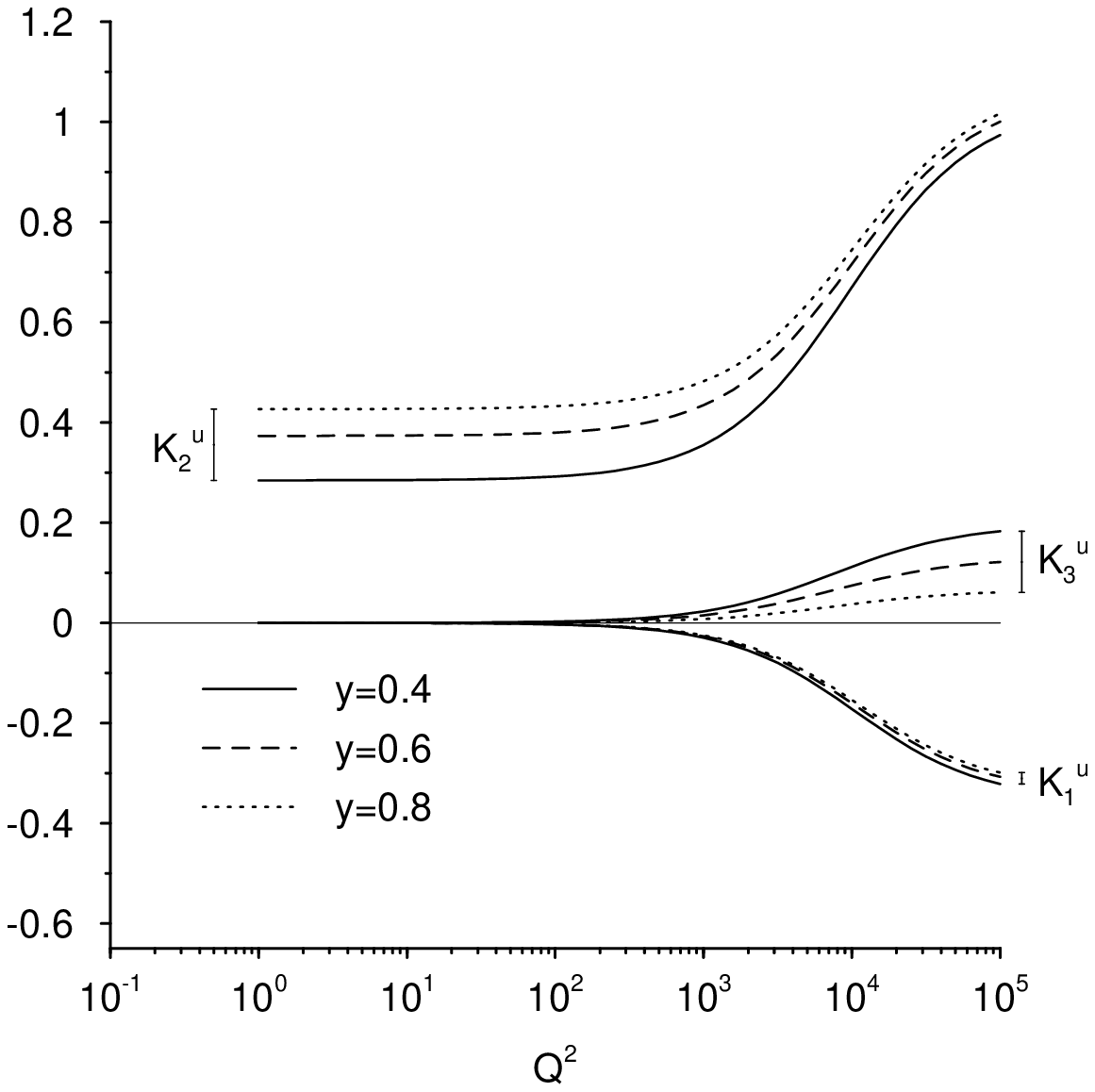, width=8.2cm}\\
\end{center} 
\caption{\label{u-coupl}
The couplings $K_i^{u}(y,Q^2)$ for three different fixed 
values of $y$: $y=0.4$ solid line, $y=0.6$ dashed line, and $y=0.8$ 
dotted line. 
On the left: sum of couplings for $\lambda_e=+1$ and $\lambda_e=-1$ 
(unpolarized beam).
On the right: difference of couplings for $\lambda_e=+1$ and $\lambda_e=-1$ 
(polarized beam).} 
\end{figure}

\begin{figure}[htb]
\begin{center}
\psfig{file=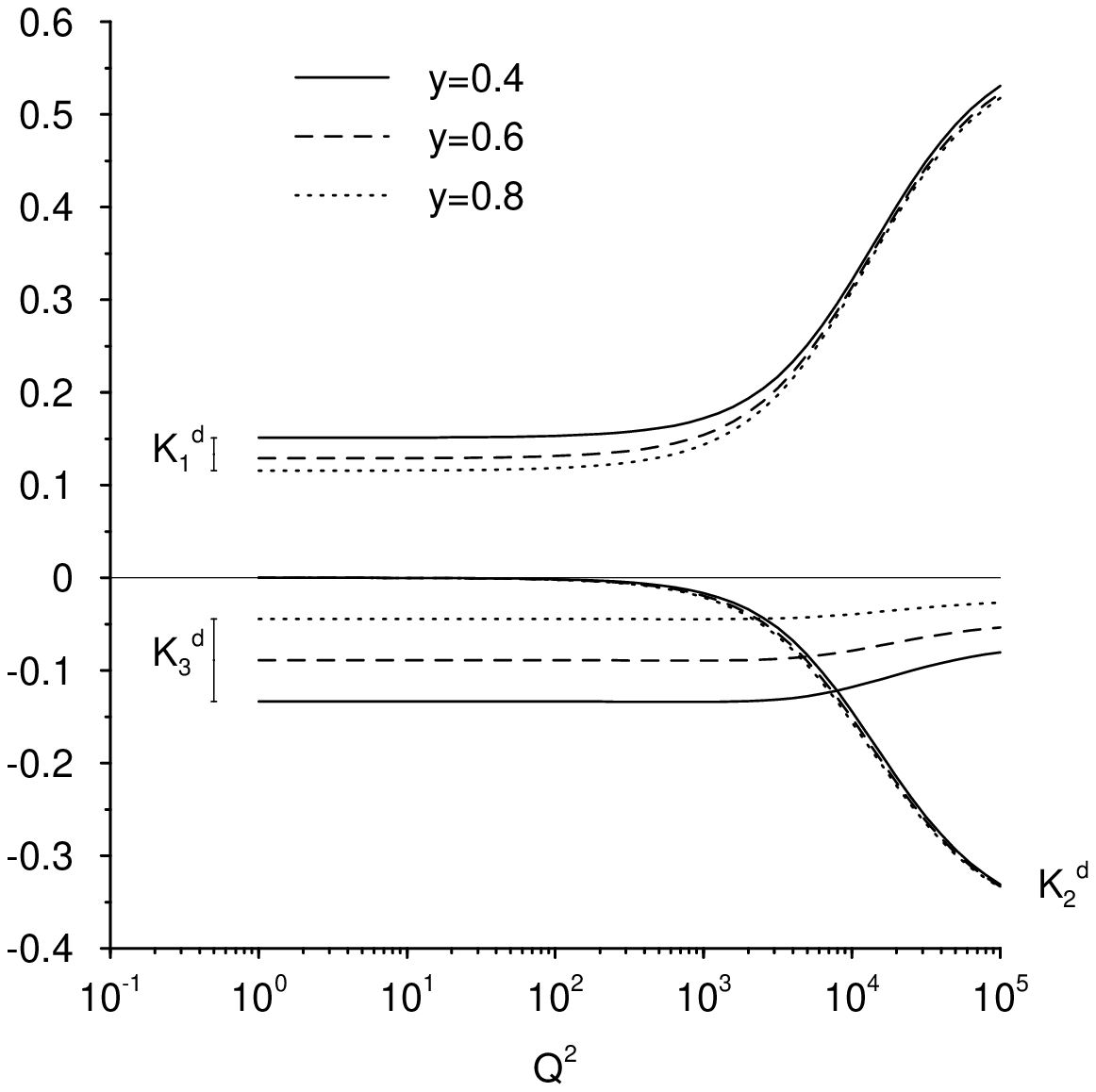, width=8.2cm}\quad
\psfig{file=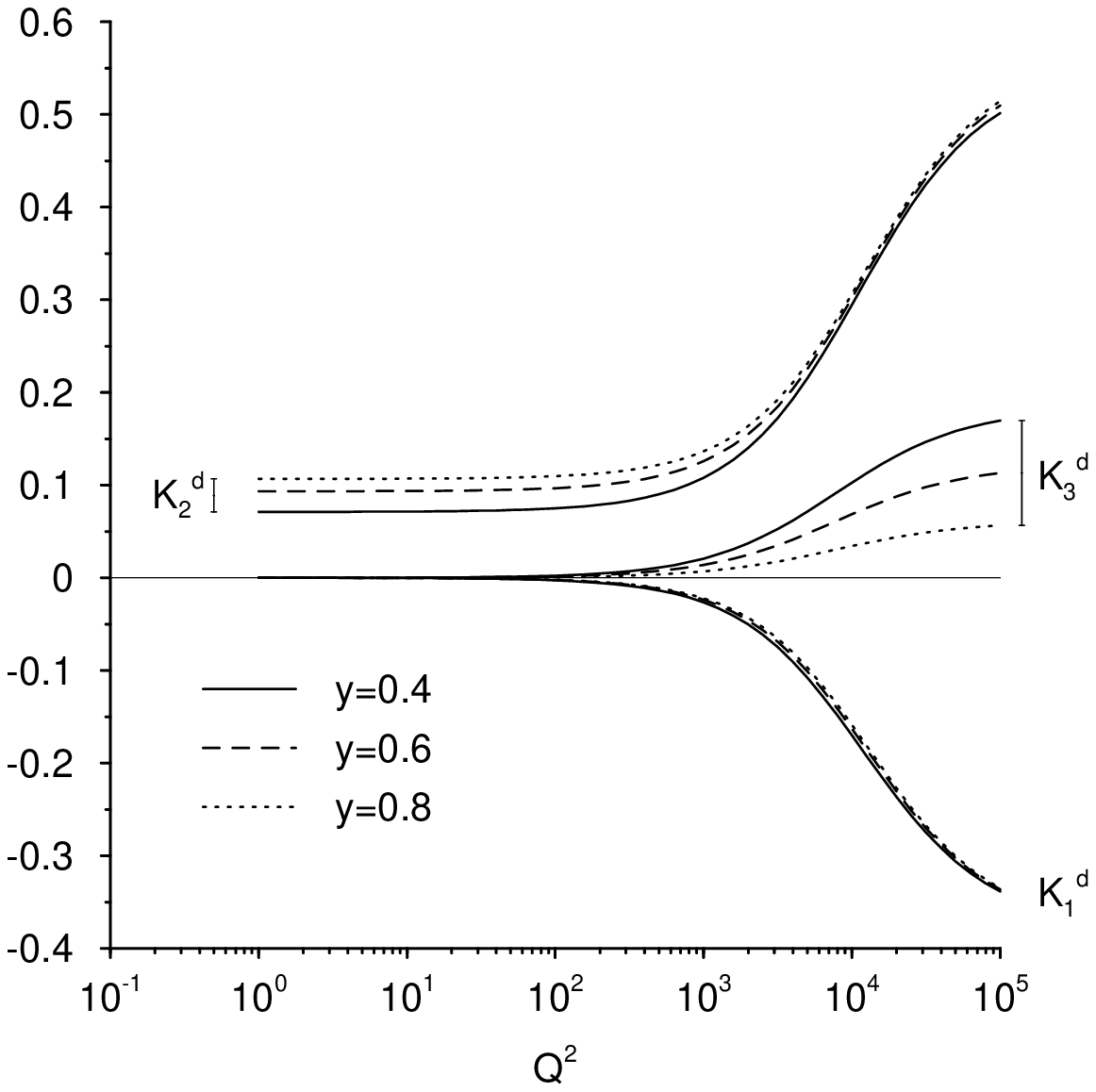, width=8.2cm}\\
\end{center} 
\caption{\label{d-coupl}
The couplings $K_i^{d}(y,Q^2)$ for three different fixed 
values of $y$: $y=0.4$ solid line, $y=0.6$ dashed line, and $y=0.8$ 
dotted line. 
On the left: sum of couplings for $\lambda_e=+1$ and $\lambda_e=-1$ 
(unpolarized beam).
On the right: difference of couplings for $\lambda_e=+1$ and $\lambda_e=-1$ 
(polarized beam).}
\end{figure}

To illustrate the kinematical regions where contributions 
from $Z$-boson exchange and interference terms become important (for
experimental data on azimuthal asymmetries at relatively 
low values of $Q^2$ cf.\ \cite{Exp}), we
plot the values of the couplings $K_i^a(y,Q^2)$ as appearing in 
the neutral current cross sections of scattering with electrons or negatively
charged muons, for different fixed values of $y$ over the range 
$1\le Q^2 \le 10^5\,{\rm GeV}^2$. Fig.\ \ref{u-coupl} shows the couplings 
for terms when the struck quark is $u$-like, and Fig.\ \ref{d-coupl} when 
it is a $d$-like quark. In both figures the two linear combinations 
$K_i^a(y,Q^2,\lambda_e=+1)+K_i^a(y,Q^2,\lambda_e=-1)$ and
$K_i^a(y,Q^2,\lambda_e=+1)-K_i^a(y,Q^2,\lambda_e=-1)$ are plotted as occurring
in scattering processes with unpolarized and polarized lepton beams,
respectively. In case one scatters with positrons or positively charged muons
one has to change the sign of $\lambda_e$, hence, the resulting plots for
unpolarized positron beams are identical to the electron scattering case and 
for polarized beams the given curves for the couplings simply change sign.  

Deviations from the $Q^2$ independent behavior\footnote{In this study we
consider only tree level, hence, we do not include logarithmic $Q^2$ behavior
due to the perturbative running of the strong and electroweak coupling
constants.  
But this will not affect the relative magnitude of the 
couplings $K_i^a$ as a
function of $Q^2$.} at low $Q^2$ indicate where interference terms are
important. Generally, effects of the weak interaction start to show up 
above $Q^2 \sim 300\,{\rm GeV}^2$ and become significant 
for $Q^2 \simorder 10^3\,{\rm GeV}^2$. It 
can be read off from the figures that asymmetries involving $K_1^a$ and 
$K_3^a$ are best measured with an unpolarized beam 
at lower values of $y$. In particular for asymmetries proportional to the 
$K_3^a$ couplings the $y$ dependence is sizable. On the other hand, 
asymmetries 
involving the $K_2^a$ couplings are best measured with a polarized 
beam, preferably at high values of $y$ (although the $y$-dependence here 
is rather moderate). Moreover, from the r.h.s.\ of the figures one sees 
that $K_2^a$ gets an 
enhancement by a factor 2 for $u$-like quarks and a factor 4 for $d$-like 
quarks at very high $Q^2$ (the same holds for the coupling $K_1^a$ for
unpolarized lepton scattering). 

Looking at the definition of $K_2^a$ in Eq.~(\ref{eq:defK2}) reveals that
azimuthal asymmetries involving these particular couplings, which are 
typical polarization measurements, can be determined even in experiments 
with an unpolarized 
beam. The axial vector coupling of the $Z$-boson provides the necessary
$\gamma_5$ structure needed to render the involved traces in the calculation
non-vanishing, which in the case of a polarized beam is provided by the
helicity terms in the lepton tensor. The observation that polarization
of quarks can be tested with unpolarized beams using electroweak interference
effects was made in the context of
electron-positron annihilation~\cite{Augustin-Renard-80}. 
The l.h.s.\ plots in 
Figs.\ \ref{u-coupl} and \ref{d-coupl}, respectively, indicate that 
azimuthal asymmetries involving $K_2^a$ couplings are accessible with 
unpolarized lepton beams for $Q^2 \simorder 10^3\,{\rm GeV}^2$. The
strength of the $K_2^a$ couplings has a small $y$ dependence. 
  
Many of the azimuthal asymmetries that arise in the above given cross
sections are difficult to measure. Nevertheless, we have presented the
complete expressions, since different terms could be accessed in different
(future) experiments. 
First of all, in order to go beyond the photon contributions one
needs a relatively high energy experiment. On the other hand, at higher
energies effects due to intrinsic transverse momentum are expected to be less
important \cite{Collins-93b}, although not power-suppressed. 
One way around the problem of having to go to very high energies is 
by studying 
semi-inclusive deep inelastic leptoproduction involving neutrinos. One can
either investigate the case of a neutrino beam scattering off a hadron (for
instance in an experiment like NOMAD at CERN) or
scattering with an electron or muon beam off a hadron with a neutrino as 
produced lepton (for instance COMPASS at CERN 
or ZEUS and H1 at DESY). In both cases no interference 
with photons occurs. On the other hand, in the case of a produced neutrino a 
new problem is that one cannot define a
lepton scattering plane as given in Fig.\ \ref{fig1} 
(one does not observe $l'$), hence
azimuthal angles cannot be defined, unless one can reconstruct the direction 
of the neutrino by the momentum imbalance \cite{Ahmed}. One could also define
azimuthal angles with respect to a transverse polarization vector of one 
initial or final state hadron, but this still limits the number of 
accessible asymmetries severely and makes the analysis much more difficult.  

In order to arrive at the expressions for the cross sections
of such charged current processes\footnote{A high $p_\sT^{}$ azimuthal 
asymmetry
in charged current semi-inclusive deep inelastic scattering arising at order
$\alpha_s$ has been studied
in Ref.\ \cite{Chay}. In Ref.\ \cite{Chay2} the same asymmetry for the neutral
current process was studied, also
at low $p_\sT^{}$, taking into account purely photon exchange. The mechanism to
include intrinsic transverse momentum is the one of \cite{Cahn} and is
therefore considered to be of higher twist.}, one can 
take $e_a=0, \, g_V^l=g_A^l=1, \, c_2^a=0$ and replace 
\beq
\chi_{\Z\Z} \to \chi_{\W\W} = \left( \frac{1}{8 \sin^2 \theta_W} \, 
\frac{Q^2}{Q^2+M_W^2} \right)^2 \;, 
\eeq
in the above given couplings $K_i^a$. In addition, one 
replaces $c_1^a=\pm c_3^a = 1$, depending on the chirality of the 
quark, since $c_1^a= (g_R^a{}^2+ g_L^a{}^2)/2$ and 
$c_3^a= (g_L^a{}^2- g_R^a{}^2)/2$. Hence, for a left-handed
quark one finds $c_1^a= c_3^a=1$ and for a right-handed
quark one finds $c_1^a= -c_3^a=1$.
This results in
\ba
K_1^{ab}(y)&=& -K_2^{ab}(y) = 4 (1-\lambda_e ) \, \chi_{\W\W} |V_{ab}|^2
\left(A(y) \pm \frac{C(y)}{2} \right) \nn\\
& = & 4 (1-\lambda_e ) \, \chi_{\W\W} |V_{ab}|^2 \, \times \Biggl\{ 
\begin{array}{cl} 1 & \quad \text{for a left-handed quark} \\ (1-y)^2 & 
\quad \text{for a right-handed quark} \end{array} \;, \label{K12-charged}\\
K_3^{ab}(y)&=& 0 \;,
\ea
where $a,b$ are the incoming and outgoing quark flavor indices, respectively, 
and $V_{ab}$ stands for the appropriate CKM matrix element. Needless to say, 
the sum over flavors in the cross section
expressions will now only run over the appropriate flavors ($u$-like or
$d$-like).  
We have neglected the lepton masses, hence helicity states 
($\lambda_e = \pm 1$) are equal to the chirality states ($R/L$). The term
$(1-\lambda_e )$ reflects the fact that the incoming lepton ($e^-,\mu^-,\nu$)
can only be lefthanded in charged exchange and the incoming anti-lepton
($e^+,\mu^+,\bar \nu$) only righthanded ($\lambda_e$ for the incoming lepton
is replaced by $-\lambda_e$ for an incoming anti-lepton). 
For example, for the elementary
process $\nu d \to e^- u$, one finds the coupling $K_1^{du} = 8 \, \chi_{\W\W}
|V_{ud}|^2 $ and for $\nu \bar u \to e^- \bar d$ one finds $K_1^{\bar u \bar
d} = 8 \, \chi_{\W\W} |V_{ud}|^2 (1-y)^2$. 
The $y$ independence of $K_1^{du}$ is explained by the fact 
that the total spin of a lefthanded lepton and a
lefthanded quark is $J=0$ such that the partonic scattering becomes isotropic
in the c.m.\ lepton scattering angle; the coupling $K_1^{\bar u \bar d}$ has a 
$(1-y)^2$ 
dependence characteristic of the $J=1$ total spin of a lefthanded neutrino and
a righthanded anti-quark (see for instance \cite{Roberts}). Note that these two
elementary processes will be accompanied by distribution and
fragmentation functions that in general are completely different in magnitude. 
The same holds for the difference between neutrino/electron versus 
anti-neutrino/positron scattering, cf.\ e.g.\ \cite{Hagiwara}.
One also has to be aware that for the charged current 
cross-sections no averaging over initial lepton polarizations has to be
performed.

The present HERA experiments ZEUS and H1 could access asymmetries for which
it is important to take into account the $Z$ boson. However, since in these
experiments no initial
polarization is present (although longitudinal lepton beam polarization will 
become disposable also for the collider experiments in the near 
future), one way to access some of the interesting asymmetries
is by focusing on $\Lambda$ production, since the spin vector of 
the $\Lambda$ can be determined from its subsequent decay. From the 
enhancement 
of the coupling $K_2$ at high $Q^2$ as apparent on the r.h.s.\ 
of Figs.\ \ref{u-coupl}
and \ref{d-coupl}, one concludes that it seems quite
promising to measure the helicity fragmentation function $G_1$ by studying 
for instance $\Lambda$ production using a polarized lepton
beam on an unpolarized target at very high $Q^2$. This option was investigated
in for instance Refs.\ \cite{Jaffe-96,Kotzinian-98,DeFlorian}. 
In this case 
one is sensitive to the 
first term in Eq.~(\ref{pol-h}), which does not involve a weight
factor. Hence, one 
can integrate over the transverse momentum of the vector boson, 
thereby deconvoluting the 
distribution and fragmentation function. The resulting cross section is
proportional to the integrated function $G_1(z)$ multiplied with the
well-known function $f_1(x)$. 
As mentioned above, from the l.h.s.\ of the 
same figures one sees that 
azimuthal asymmetries involving $K_2^a$ couplings are also accessible with 
unpolarized lepton beams for $Q^2 \simorder 10^3\,{\rm GeV}^2$, thus 
allowing for a measurement of $G_1$ with {\em unpolarized\/} 
lepton and proton beams. This is the semi-inclusive deep inelastic scattering 
analogue
of the proposed measurement of $G_1$ in $e^+ e^- \to \Lambda^\uparrow X$
\cite{Burkardt}. One could also exploit charged current exchange by using 
a neutrino (or anti-neutrino) beam like for instance in $\nu p \to \mu^-
\Lambda^\uparrow X$ \cite{Kotzinian-98,Ma-Soffer}, but we 
would like to emphasize that one can use the {\em neutral current\/} 
exchange process 
without the need for lepton beam polarization, 
i.e.\ $\ell H \to \ell^\prime \Lambda^\uparrow X$. 

A polarized proton beam, for instance at the proposed polarized HERA collider
\cite{Hughes-99}, 
would give even more opportunities to measure the different asymmetries
presented here, i.e.\ in principle those given in 
Eqs.\ (\ref{reducedO}) and (\ref{pol-Hh}), cf.\ also Ref.\ \cite{Maul}. A
remark that is relevant here is that measuring the transversity functions
$h_1$ and $H_1$ via the $\cos(\phi_{S}+\phi_{S_h})$ term in 
Eq.\ (\ref{pol-Hh}) cannot be done in the charged current 
exchange case (e.g.\ via 
$e^- p^\uparrow \to \nu \Lambda^\uparrow X$), since $K_3^{ab}(y)=0$. The same
holds for any other chiral-odd function. But they 
can of course be accessed in the
neutral current processes and at very high energies lepton beam polarization 
can even be of assistance.   

In conclusion, we have presented the leading order unpolarized and 
polarized cross sections in electroweak semi-inclusive deep inelastic 
leptoproduction. We have discussed the
present and future possibilities for experimental investigation of some of the
asymmetries presented here. In particular,  
the opportunities offered by neutral and charged current 
processes were contrasted and the optimal kinematic regions (in $y$ and $Q^2$)
for which one might expect certain asymmetries to be measurable were studied. 
We have observed that one can measure the helicity fragmentation function 
$G_1$ by $\Lambda$ production in the neutral current exchange process with 
both lepton and proton beams unpolarized. Also, we have noted that 
the transversity distribution and fragmentation functions cannot be
measured in charged current exchange semi-inclusive leptoproduction.  

\acknowledgments 
D.B.\ thanks Matthias Grosse-Perdekamp for useful discussions. 
R.J.\ is supported by the European TMR program ERB FMRX-CT96-0008 of the
network Hadronic Physics with High Energy Electromagnetic Probes.

\end{document}